\documentclass[usenatbib,useAMS]{mn2e}
\usepackage{epsfig}
\usepackage{times}
\usepackage{amsmath}

\usepackage{verbatim}

\def\d{\mbox{d}}
\newcommand{\simgt}%
        {\,\hbox{\lower0.6ex\hbox{$\sim$}\llap{\raise0.6ex\hbox{$>$}}}\,}
\newcommand{\simlt}%
        {\,\hbox{\lower0.6ex\hbox{$\sim$}\llap{\raise0.6ex\hbox{$<$}}}\,}

\title[]{Multi-dimensional modelling of X-ray spectra for 
AGN accretion-disk outflows}
\author[Sim et al.]{S. A. Sim$^1$, K. S. Long$^2$, L. Miller$^3$, T. J. Turner$^{4,5}$\\
$^{1}$Max-Planck-Institut f\"{u}r Astrophysik,
Karl-Schwarzschildstr. 1, 85748 Garching, Germany\\
$^{2}$Space Telescope Science Institute, 3700 San Martin Drive,
Baltimore, MD 21218, U.S.A\\
$^{3}$Dept. of Physics, University of Oxford, Denys Wilkinson
Building, Keble Road, Oxford OX1 3RH, U.K.\\
$^{4}$Dept. of Physics, University of Maryland Baltimore County, 1000 Hilltop Circle, Baltimore, MD 21250, U.S.A\\
$^{5}$Astrophysics Science Division,
NASA/GSFC, Greenbelt, MD 20771, U.S.A\\
}

\date{\today}

\pagerange{\pageref{firstpage}--\pageref{lastpage}}
\pubyear{}
\begin{document}
\maketitle
\label{firstpage}

\begin{abstract}
We use a multi-dimensional Monte Carlo code to compute
X-ray spectra for a variety of active galactic nucleus (AGN) 
disk-wind outflow geometries.
We focus on the formation of blue-shifted absorption features in the
Fe~K band and show that line features similar to those which have been
reported in observations are often
produced for lines-of-sight through disk-wind geometries.
We also discuss the formation of other spectral features in highly
ionized outflows. In particular we show that, for sufficiently high
wind densities, moderately strong Fe~K emission lines can form 
and that electron scattering in the flow may cause these lines to develop
extended red wings. We
illustrate the potential relevance of such models to the
interpretation of real X-ray data by comparison with observations of
a well-known AGN, Mrk~766.
\end{abstract}

\begin{keywords}
radiative transfer --  methods: numerical -- galaxies: active --
X-rays: galaxies
\end{keywords}

\section{Introduction}
\label{sect_intro}

The study of active galactic nuclei (AGN) is an important topic in
contemporary astrophysics. These objects are interesting in their own
right, allowing us to study the extreme physics of accretion in the
vicinity of a supermassive black hole. Furthermore,
numerical simulations suggest that AGN likely have a critical role 
in the formation 
and evolution of galaxies, highlighting the need to understand 
how these objects accrete
matter, grow and feedback energy to their surroundings 
(e.g. \citealt{croton06}).

Since AGN are bright X-ray sources, they have been popular targets for
almost all X-ray observatories. One of the most interesting results
obtained thanks to the high sensitivity of the current generation of X-ray missions (specifically
{\it XMM-Newton} [\citealt{jansen01}], {\it Chandra}
[\citealt{weisskopf02}]  and, 
more recently, {\it Suzaku} [\citealt{mitsuda07}])
has been the detection of narrow absorption features in the 
2 -- 10~keV band of several bright AGN (see, e.g. \citealt{pounds03,
reeves04,risaliti05,young05,turner07,braito07}). 
Particularly for features in the Fe~K region of the spectrum, the
observed energies and strengths of these lines suggest identification
with very highly ionized species (e.g. Fe~{\sc xxv} and {\sc xxvi}) in
very fast (up to $\sim 0.1$c) outflows.
Blueshifted absorption lines are well-known from observations of AGN in other
wavebands and models explaining such phenomena in terms of winds 
have been developed (see e.g. \citealt{murray95}, 
\citealt{elvis00}, \citealt{proga04}). However, the new X-ray data clearly
suggest a component of very highly ionized fast outflowing plasma, the
relationship of which to the less ionized material observable in
ultraviolet (uv) or softer X-ray wavebands remains unclear. In particular,
moderately ionized 
outflows identified in soft X-ray spectra generally have somewhat
lower velocity ($\simlt 1000$~km~s$^{-1}$; \citealt{blustin05}, 
\citealt{mckernan07}) while
uv spectra can show evidence of either low or high velocity absorption
(e.g. \citealt{elvis00}).

As
observational evidence in support of the phenomenon has accumulated,
several theoretical studies of the 
physical properties of highly ionized AGN outflows have been made.
\citet{pounds03} and \citet{king03} discussed the
blueshifted absorption features in PG1211+143 in terms of a
conical outflow subtending a large solid angle. They suggested that
such a flow might be driven by continuum radiation pressure and that
emission from such a flow might be responsible for the big blue bump
which dominates the bolometric output of
PG1211+143. However, \citet{everett04} considered flows driven by
continuum radiation pressure in greater detail and concluded that,
while this mechanism could work in principle, the resulting outflows
were unlikely to produce spectral signatures as strong as those
reported by \citet{pounds03}.

\citet{sim05b} undertook a two-dimensional (2D) Monte Carlo (MC) radiative
transfer study of parameterised conical outflow models and concluded
that, for suitable column densities, viewing down such flows
could account for the observed blueshifted, narrow absorption lines in
PG1211+143. However, that study was limited to consideration of only
the simplest conical geometry and did not consider the effect of
either rotation or off-axis lines-of-sight on the spectrum.

Using a chained version of the 1D XSTAR code \citep{kallman01},
\citet{schurch07,schurch08} have computed spectra for columns of
outflowing gas with a variety of density and velocity profiles. They
demonstrated that outflows can imprint a wide variety of features on
X-ray spectra, depending on the outflow conditions and that very high
outflow velocities would be required if absorption in outflows is
to explain soft excesses in AGN spectra. More recently, a 
similar approach to that adopted by \citet{schurch07} has been used
for the calculation of transmission spectra in dynamical models of AGN
outflows by \citet{dorodnitsyn08}.

In this paper, we extend the study of \citet{sim05b} to incorporate a
more realistic and more versatile description of a disk wind 
geometry including both rotation and off-axis lines-of-sight. 
We retain the use of the MC method owing to its versatility
for multi-dimensional radiative transfer. 
Our method is complementary
to that of \citet{schurch07} since we account for geometric effects
directly (e.g. scattering of radiation between lines-of-sight in
multi-dimensional outflow geometries) but make some simplifications in
the treatment of atomic processes.

We focus on highly ionized winds and the interpretation of observable 
features in the Fe~K band. 
Although other spectral
lines are expected to form in outflows, Fe~K$\alpha$ absorption is
the most relevant for the interpretation of observed spectra since it
is the clearest signature of a highly ionized flow: although more
abundant, the lighter elements are
fully ionized more easily than Fe so that any spectral features they
imprint are weaker. Furthermore, identification
of features in the Fe~K energy band is relatively secure since only K
shell transitions of heavy ions are expected at energies $\simgt
5$~keV. At lower energies, where K$\alpha$
transitions of light or intermediate mass elements might arise, line
identification is more ambiguous, particularly if large velocity shifts
are considered. 

Although we regard Fe~K$\alpha$ absorption features as the best
diagnostic for a highly ionized flow, we will also investigate the
formation of emission
components in the Fe~K arising from line
scattering or recombination in the outflow. Such features are of
particular interest since they may explain possible P-Cygni-like line
profiles (e.g. \citealt{done07}, \citealt{turner08}) and may affect
the interpretation of
Fe emission features commonly associated with disk reflection {
(see e.g. 
\citealt{laming04}, \citealt{laurent07}).}

We begin in Section~\ref{sect_model} by defining the class of outflow
models which we will consider. In Section~\ref{sect_RT} we describe
our radiative transfer method and discuss the adopted atomic data. We
present a sample calculation for one model in
Section~\ref{sect_example} and then extend our discussion to a a grid
of outflow models in Section~\ref{sect_grid}. The implications of our
models for observations of Fe~K$\alpha$ absorption are described in
Section~\ref{sect_obs} and for other spectral features in
Section~\ref{sect_beyondka}. 
In Section~\ref{sect_mrk766} we illustrate the value of our models by
comparing them in detail with observations of a well-known AGN, Mrk~766.
We draw conclusions and discuss further
work in Section~\ref{sect_conc}.

\section{Model}
\label{sect_model}

Our radiative transfer calculations were
performed using a simply-parameterised model for an outflow launched
from an accretion disk around a supermassive black hole. In this
section, we describe the properties of the model and the parameters
which must be specified to define a particular realisation of the
model.

\subsection{Geometry}

\begin{figure}
\epsfig{file=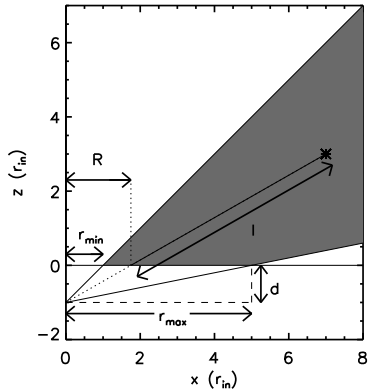, width=8cm}
\caption{
The geometrical construction used to define the wind 
(only the positive $xz$-plane
is shown -- the wind is symmetric under both reflection in the $xy$-plane and rotation about the $z$-axis).
The region occupied by the wind is shaded. The three parameters which
determine the geometry ($d$, $r_{\mbox{\scriptsize min}}$, $r_{\mbox{\scriptsize max}}$) are indicated. 
The star symbol represents the point in the wind which is
specified by the wind coordinates $R$ and $l$.
}
\label{fig_geo}
\end{figure}

We adopt a standard disk wind geometry, namely the ``displaced
dipole'' model of \citet{knigge95}. This
geometry has been 
adopted in radiative transfer studies of accretion disk winds for a
variety of systems including cataclysmic variables \citep{knigge95,
  long02} and massive young stellar objects \citep{sim05}.
The geometry is illustrated in Fig.~\ref{fig_geo} and is defined by
the following three parameters (each of which is marked in the
figure):

\begin{enumerate}

\item {$d$}, the distance of the focus point below the origin

\item {$r_{\mbox{\scriptsize min}}$}, the distance from the origin to the inner edge of the wind in the $xy$-plane

\item {$r_{\mbox{\scriptsize max}}$}, the distance from the origin to the outer edge of the wind in the $xy$-plane

\end{enumerate}

The accretion disk is assumed to lie in the $xy$-plane.
The wind is symmetric under rotation about the $z$-axis and under reflection in the $xy$-plane.

\subsection{Velocity law in the wind}

The velocity is specified at every point in the wind following the 
parameterisation of \citet{knigge95} (see also \citealt{long02}).
Although this velocity law is not based on self-consistent
hydrodynamical outflow models (cf. \citealt{dorodnitsyn08}),
it provides a simple and reasonably flexible description of possible
steady-state flows
which we use for our exploratory radiative transfer
simulations. 

\subsubsection{Rotation}

The wind rotates about the $z$-axis. The rotational velocity is obtained
by assuming that parcels of matter conserve specific angular momentum
about the $z$-axis as they flow outwards. The angular momentum at the
base of a streamline is set at the Keplerian value for the
radius at which the streamline crosses the $xy$-plane. Thus the
rotational velocity is determined only by the choice of wind geometry
(see above) and the mass of the central object, $M_{\mbox{\scriptsize bh}}$.

\subsubsection{Outflow}

The outflow velocity points directly away from the focus point of the
wind.
Its magnitude is given by

\begin{equation}
v_{l} = v_{0} + (v_{\infty} - v_{0}) \left( {1 - \frac{R_{v}}{R_{v} + l}} \right)^{\beta}
\end{equation}
where $l$ is the distance along the outflow streamline (see
Fig.~\ref{fig_geo}), $R_{v}$ is 
an acceleration length parameter and the exponent, $\beta$, sets the
rate of acceleration. 
The terminal velocity, $v_{\infty}$, is
specified
as a multiple $f_{v}$ of the escape speed from the base of the streamline

\begin{equation}
v_{\infty} = f_{v} \sqrt{ \frac{2 G M_{\mbox{\scriptsize bh}}}{R}}
\end{equation}
where $R$ is the radius at the base of the streamline (see Fig.~\ref{fig_geo}).

\subsection{Mass density}

The wind is assumed to be smooth and in a steady state flow;
extension of this study to clumpy flows will be the topic of a
subsequent investigation. The total mass-loss rate ($\dot{M}$) of the
wind is treated as a parameter and the mass loading of specific
outflow streamlines is determined by parameterising the 
mass-loss rate per unit surface area as a function of $R$ via

\begin{equation}
\frac{d \dot{m} }{d A} \propto R^{k}
\end{equation}
subject to the constraint that

\begin{equation}
4\pi \int_{r_{\mbox{\scriptsize min}}}^{r_{\mbox{\scriptsize max}}} \frac{d \dot{m} }{d A} \; R \; d R = \dot{M} \; \; .
\end{equation}
The mass density at a point in the wind is then obtained by combining
the specific mass-loss rate on local outflow streamlines with the outflow velocity, assuming a stationary flow.

\subsection{Electron temperature}

The electron temperature $T_{e}$ of the outflowing material must be specified
since it has a role to play in determining the ionization/excitation
state of the plasma (see below). Ultimately, it should be
calculated by consideration of all relevant heating and cooling
processes. But the treatment of all such processes --
heating and cooling via lines, bound-free continua, free-free
processes and (inverse-)Compton scattering of photons at all energies -- 
would require more detailed atomic
physics, simulations of a much wider range of photon energies
and the self-consistent
consideration of heating by
radiation from other sources 
(e.g. radiation emitted or reflected by
the accretion disk or surrounding structures).
Such issues go beyond the scope of this study and so, as 
in \citet{sim05b}, $T_e$ is assumed
to be uniform and is treated as an input parameter.

\subsection{Atomic level populations}

In order to keep the computational requirements of our
multi-dimensional MC calculations tractable, we have
not attempted a complete non-LTE calculation, but
have instead employed approximate formulae for
ionization and excitation.  Our approach is similar
to the standard modified nebular approximation,
which has been used in a variety of other
astrophysical calculations 
(e.g. stellar winds [\citealt{abbott85}, \citealt{lucy93},
\citealt{vink99}], supernovae [\citealt{mazzali93}] and disk winds from
cataclysmic variables [\citealt{long02}]). 
However, as
discussed in Section~\ref{sect_radfld}, we have modified
standard approximations to reflect the fact that
X-ray spectra of AGN are better approximated by a
power law than a blackbody.

\subsubsection{Ionization}

For simplicity, we assume ionization equilibrium in which 
the dominant ionization process for all
the ions we consider is photoionization from the ground
state. Furthermore, we neglect recombinations from excited states such
that the condition of ionization equilibrium can be expressed:

\begin{equation}
n_{i,0} \gamma_{i,0} = n_{e}
n_{i+1,0} \sum_l \alpha_{i+1,0 \rightarrow l}
\end{equation}
where $n_{i,0}$ is the atomic level population of an ionization
state $i$ in its ground state (denoted $0$), $n_{e}$ is the free
electron density, $\gamma_{i,0}$ is the photoionization rate
coefficient from the ground state of ion $i$ and 
$\alpha_{{i+1,0 \rightarrow l}}$ is the recombination rate coefficient
from the ground state of ionization state $i+1$ to level $l$ of
ionization state $i$. The summation runs over all states of the lower
ion.

This can be rewritten as

\begin{equation}
\frac{n_{i+1,0} n_{e}}{n_{i,0}} = \zeta(T_{e}) S_{i}(T_{e}, J_{\nu}) \Phi_{0,0,i}(T_{e})
\label{eqn_ionbal}
\end{equation}
where $\Phi_{0,0,i}(T_{e})$ is the population ratio ${n_{i+1,0}
n_{e}}/{n_{i,0}}$ evaluated for LTE at the electron temperature
($T_{e}$) and

\begin{equation}
\zeta = \frac{\alpha_{i+1,0 \rightarrow 0}}{\sum_l \alpha_{i+1,0
    \rightarrow l}}
\end{equation}
is the fraction of recombinations that go directly to the ground state.

\begin{equation}
S_{i}(T_{e}, J) = \frac{\gamma_{i,0 \rightarrow 0}}{\gamma_{i,0 \rightarrow 0}^{LTE}}
= \int_{\nu_0}^{\infty} a_{\nu} J_{\nu} \nu^{-1}\; \mbox{d} \nu / \int_{\nu_0}^{\infty} a_{\nu} B_{\nu}(T_{e}) \nu^{-1}\; \mbox{d} \nu
\end{equation}
is the ratio of the true photoionization rate determined from the mean
intensity ($J_{\nu}$) and that which would be obtained in LTE at
$T_e$. $a_{\nu}$ is the ground state 
photoionization cross-section for frequency
$\nu$, $B_{\nu}$ is the Planck function and the integrations run from the
threshold frequency, $\nu_{0}$, to infinity.

Equation~\ref{eqn_ionbal} is used for all the ionization calculations
in this paper. Since $\Phi_{0,0,i}$ and $\zeta$ depend only on the
adopted electron temperature these can be
evaluated prior to the MC radiative transfer simulation. $S_{i}$, however,
depends on the radiation field $J_{\nu}$ and so it is computed from the
radiation field parameterisation and iterated
(see Section~\ref{sect_radfld}).

\subsubsection{Excitation}

Since the majority of the atomic processes with which we are concerned
are associated with ground state absorption, the treatment of excited
level populations is relatively unimportant and therefore very
approximate. 
For all excited states, we
adopt a two-level radiation-dominated Sobolev approximation

\begin{equation}
\frac{n_l}{n_0} = \frac{g_l}{g_0} \left( {{\frac{2 h \nu_{l}^{3}
\bar{\beta}}{\bar{J_{\nu}} c^2} + 1} } \right)^{-1}
\label{eqn_excit}
\end{equation}
where $h \nu_{l}$ is the excitation energy of state $l$ above the ground
state, $g_l$ is the statistical weight of level $l$, $\bar{\beta}$ is the
angle-averaged Sobolev escape probability for the transition from
level $l$ to the ground state and $\bar{J_{\nu}}$ is the mean intensity in
the transition.

\section{Radiation transport simulations}
\label{sect_RT}

The propagation of radiation through the model is followed via a MC
simulation in which the quanta are indivisible packets of radiative
energy. 
The principles of the method have been developed and described
elsewhere (e.g. \citealt{abbott85}, \citealt{lucy93},
\citealt{mazzali93}, \citealt{lucy02}, \citealt{lucy03}) 
so only those points of specific relevance to
the procedure used in this study are given below.

\subsection{Computational grid}

For the radiative transfer calculations, the wind properties are first
discretised onto a grid in the natural wind parameters $(R,
l)$. Typically a 100 x 100 grid is used. For convenience when
propagating the packets, this grid of wind properties is then mapped to
an underlying
three-dimensional (3D) Cartesian grid, typically 100x100x100. 
Each Cartesian cell which lies inside the wind 
is assigned the density, ionization/excitation state and radiation
field properties of the closest point in the grid of wind properties
while those Cartesian grid cells which lie outside the wind are
assumed to be empty.

\subsection{Initialisation of packets}

Since our primary objective is to simulate spectra in the keV energy range, we
initially create packets between 0.1 and 40 keV. A power-law
distribution of packet energies is assumed such that the input
spectrum $n(E)$
varies with photon energy ($E$) according to

\begin{equation}
n(E)[\mbox{photons~s$^{-1}$~keV$^{-1}$}] \propto E^{-\Gamma} \;.
\label{eqn_inrad}
\end{equation}
This is an adequate
first-order description of the observed X-ray spectra of AGN in the 2
-- 10 keV range with which we are primarily concerned. 
The power-law
index ($\Gamma$) is as an input parameter for our model.

It is assumed that the primary X-ray source is compact and roughly
appears as a point source as seen from the outflowing gas. Thus, 
the packets are launched from the coordinate origin. Their initial
directions are chosen isotropically.

\subsection{Propagation of packets}

Once launched, the packets  propagate until they reach the
outer boundary of the computational domain. As they propagate, the
packets can interact with the outflowing plasma, thereby changing both
their direction of propagation and their observer-frame
frequency. In all interactions, energy conservation is
strictly enforced such that, in a local co-moving frame, there is no
net gain or loss of radiative energy. The particular types of packet
interaction which may occur during the simulation are summarised
below.

\subsubsection{Compton scattering}

By far the most common interaction is Compton scattering by free
electrons. This process is treated using the method outlined by
\citet{lucy05}. In all our simulations, we assume the electron
temperature is small enough that inverse Compton scattering may be
neglected in the co-moving frame.

\subsubsection{Photoabsorption}

Bound-free continua can absorb any photons with frequency above their
edges -- the cross-sections are obtained from the atomic
data sources described in Section~\ref{sect_atomicdata}. During the MC
simulations, only bound-free absorption by ionic ground states is
included.

\subsubsection{Bound-bound absorption}

Atomic lines can absorb photons that come into resonance with
them. Their optical depths are computed in the Sobolev approximation
using the velocity gradient and the appropriate level populations
obtained from the ionization/excitation formulae. 

\subsubsection{Re-emission}
\label{sect_reemiss}

Following photoabsorption or bound-bound absorption, the macro atom
scheme devised by \citet{lucy02,lucy03} is used to determine the
subsequent re-emission frequency of the packet. 
This approach naturally incorporates both resonance scattering and line
emission following recombination, the two processes which are of
greatest interest for outflow features in AGN X-ray spectra.
The particular macro atom scheme adopted allows for 
bound-bound radiative and electron collisional processes in both
internal state changes and deactivation. Bound-free recombination
processes are also included but, in accordance with the assumption
made for Equation~\ref{eqn_ionbal}, photoionization from excited
states is neglected.

Packets emitted by macro atoms via bound-free continua are radiated
isotropically. For bound-bound transitions, the direction of
re-emission is determined by constructing a 2D 
grid of Sobolev escape probabilities as a function of polar and
azimuthal angle and sampling this distribution to choose
a new packet trajectory.

\subsubsection{$k$-packets}

When the outcome of an interaction is a conversion of a radiative packet 
to a packets of thermal energy
($k$-packets in the nomenclature of \citealt{lucy03}), we
assume that the subsequent elimination of the $k$-packets
predominantly leads to emission at lower photon energies such that these
packets are lost to the hard X-ray energy band.

\subsection{Monte Carlo estimators and the radiation field}
\label{sect_radfld}

\subsubsection{Parameterisation of the radiation field}

The treatment of ionization requires that the radiation field be known
locally. Since the wind may be optically thick,
the radiation field is computed from the behaviour of the MC
packets using volume based estimators. 
In principle, it is possible to describe an arbitrary radiation field
using MC estimators for the mean intensity in narrow frequency
bins. However, this requires very large numbers of MC quanta which
becomes restrictive for the exploration of model parameter
spaces. Therefore, we follow previous MC studies in adopting a
physically motivated parameterisation of the radiation field.

This method has been described
by \citet{lucy99, lucy03, lucy05} and used in various other studies
(e.g. \citealt{mazzali93,long02}). 
In these previous studies, it was assumed that the true
radiation field was black-body in character and was therefore
usually
parameterised by a radiation temperature and a dilution factor. Here,
however, we are concerned with a radiation field which is
non-thermal in character and so parameterisation based on a power-law is
adopted:

\begin{equation}
J_{\nu}(r) = W(r) \nu^{\alpha(r)}
\end{equation}
where both the parameters $W$ and $\alpha$ are functions of
position. 
Given our choice of input radiation field (equation~\ref{eqn_inrad}),
if the wind is optically thin, we would expect to have
$\alpha(r) = 1 - \Gamma$ at all points. However, for optically thick
winds, we expect variation in $\alpha$.
Using this description, the $S_{i}$-factors in the
ionization equation can be
expressed as functions of only $T_{e}$ and the local values of $W$ and $\alpha$

\begin{equation}
S_{i}(T_{e}, W, \alpha) = W \int_{\nu_0}^{\infty} a_{\nu} \nu^{\alpha-1}\;
\mbox{d} \nu / \int_{\nu_0}^{\infty} a_{\nu} B_{\nu}(T_{e}) \nu^{-1} \;
\mbox{d} \nu \; .
\end{equation}

This allows the ionization state of the gas to be uniquely determined from
$T_{e}$, the local mass-density and the radiation field parameter pair 
[$W$, $\alpha$] only.

To specify the excitation state via equation~\ref{eqn_excit}, we also
need to know $\bar{J_{\nu}}$ for ground state transitions. Again, exact MC
estimators for $\bar{J_{\nu}}$ in all transitions can be constructed but
recording and storing them for large numbers of transitions in
multi-dimensional grids becomes expensive in terms of both processor
time and memory allocation. $\bar{J_{\nu}}$ can be expressed as

\begin{equation}
\bar{J_{\nu}} = \frac{1}{4\pi} \int_{0}^{2\pi} \int_{-1}^{1} I_{\nu,\mu,\phi} \beta_{\mu,\phi} \; \d{\mu}
\; \d\phi
\end{equation}
where $\cos^{-1} \mu$ and $\phi$ are spherical polar angles,
$I(\nu,\mu,\phi)$ is specific intensity and $\beta(\mu,\phi)$ is the
Sobolev escape probability for the transition. For the specific
case of an homologous flow, $\beta$ becomes independent of direction
but, in general, it is a function of both $\mu$ and $\phi$. In the
interests of computational expediency, however, we neglect this
dependency and replace $\beta_{\mu,\phi}$ with its angle-averaged
value, $\bar{\beta}$. This simplification allows $\bar{J_{\nu}}$ to be
expressed as

\begin{equation}
\bar{J_{\nu}} = \frac{1}{4\pi} \bar{\beta} \int_{0}^{2\pi} \int_{-1}^{+1} I_{\nu,\mu,\phi} \; \d{\mu}
\; \d\phi = \bar{\beta} J_{\nu} = W \nu^{\alpha} \bar{\beta} \; ,
\end{equation}
adopting our parameterisation of the radiation field. This simplifies
equation~\ref{eqn_excit} to give our final excitation formula

\begin{equation}
\frac{n_l}{n_0} = \frac{g_l}{g_0} \left( {{\frac{2 h}{W \nu_{l}^{\alpha-3} c^2} + 1} } \right)^{-1}
\end{equation}
where $h \nu_{l}$ is the excitation energy of state $l$ above the ion
ground state. This treatment of excitation avoids the need to specify
any radiation field parameters beyond those used in the ionization
formula and therefore places no further computational demands on the
MC simulations. It is very crude but is not of critical
importance to our study of highly ionized outflows since the
excitation state is low enough that the
ground state populations of H- and He-like species are dominant. The
most important failing of this treatment will be for the 
metastable triplet states of He-like ions. Thus improvements to the
treatment of excitation will be necessary in order to extend the code
for applicability to less ionized flows and softer wavebands.

\subsubsection{Monte Carlo estimators}

The
values of $W$ and $\alpha$ are obtained by recording two MC estimators
per 
grid cell:

\begin{equation}
E^{(1)}_{n} = \sum_{\mbox{\scriptsize paths in $n$}}
\epsilon_{\mbox{\scriptsize cmf}} \; \mbox{d}s
\end{equation}

\begin{equation}
E^{(2)}_{n} = \sum_{\mbox{\scriptsize paths in $n$}} \epsilon_{\mbox{\scriptsize cmf}} \nu_{\mbox{\scriptsize cmf}}  \; \mbox{d}s
\end{equation}
where 
the summation runs over all quanta trajectories which lie inside the
grid cell of interest (denoted by $n$) and in which the packet
frequency lies in the regime of interest ($\nu_{\mbox{\scriptsize min}} < \nu < \nu_{\mbox{\scriptsize max}}$).
In these equations, 
$\mbox{d}s$ is the length of a Monte Carlo quanta trajectory length, 
$\epsilon_{\mbox{\scriptsize cmf}}$ is the comoving frame packet energy and
$\nu_{\mbox{\scriptsize cmf}}$ is the comoving frame packet frequency.

Values for $E^{(1)}_{n}$ and $E^{(2)}_{n}$ are recorded during the
Monte Carlo simulation of all the packets. At the end of the
simulation, they are used to obtain values of $\alpha_{n}$ and $W_{n}$ 
in each grid cell as follows.

The ratio $E^{(2)}_{n} / E^{(1)}_{n}$ gives the mean frequency. Since
we know the frequency interval being simulated ($\nu_{\mbox{\scriptsize min}} < \nu <
\nu_{\mbox{\scriptsize max}}$), 
this ratio is directly linked to the power-law index
via:

\begin{equation}
\frac{E^{(2)}_{n}}{E^{(1)}_{n}} = \frac{\alpha_n + 1}{\alpha_n + 2} 
\frac{\nu_{\mbox{\scriptsize max}}^{\alpha_n + 2} - \nu_{\mbox{\scriptsize min}}^{\alpha_n + 2}}
{\nu_{\mbox{\scriptsize max}}^{\alpha_n + 1} - \nu_{\mbox{\scriptsize min}}^{\alpha_n + 1}}
\end{equation}
which can be numerically solved for $\alpha$. 

Once $\alpha_n$ is known, $W_n$ can be obtained by normalizing
$E^{(1)}_{n}$:

\begin{equation}
W_{n}
= 
\frac{{E^{(1)}_{n}}}{4 \pi V_{n} \Delta t}
\frac{\alpha_n+1}{\nu_{\mbox{\scriptsize max}}^{\alpha_n + 1} - \nu_{\mbox{\scriptsize min}}^{\alpha_n +
1}}
\end{equation}
where $V_{n}$ is the volume of grid cell $n$ and $\Delta t$ is the
effective time interval represented by the Monte Carlo simulation.

\subsubsection{Iteration}

The formulae given above allow values of $W_{n}$ and $\alpha_{n}$ to
be obtained from the behaviour of the MC quanta. However,
since the values of these parameters also affect the MC
simulation (primarily owing to their importance in determining the
ionization state), these parameters must be iterated to reach a
converged parameterisation of the radiation field.

In practice, the radiation field parameters are found to converge very
rapidly, often requiring only one iteration step.

\subsection{Atomic data}
\label{sect_atomicdata}

Table~\ref{tab_atoms} lists the elements and ionization stages that are
included in the radiative transfer calculations; for all the light
elements, only the highest three ionization stages are included while
for Fe and Ni the five highest stages are treated.

\begin{table}
\caption{Elements and ions which are included in the radiative transfer
calculations.}
\label{tab_atoms}
\begin{tabular}{ccccc}
Element & Ions &~~~~~~~~ & Element & Ions\\ \hline
C & {\sc v -- vii} & &S & {\sc xv -- xvii}\\
N & {\sc vi -- viii} & &Ar &{\sc xvii -- xix}\\
O & {\sc vii -- ix} & &Ca &{\sc xix -- xxi}\\
Ne & {\sc ix -- xi} & &Fe & {\sc xxiii -- xxvii}\\
Mg & {\sc xi -- xiii} & &Ni & {\sc xxv -- xxix}\\
Si & {\sc xiii -- xv} & & \\ \hline
\end{tabular}
\end{table}

The solar element abundances of \citet{asplund05} are adopted.

For all elements other than Fe, atomic models, bound-bound oscillator
strengths and electron collision rates were extracted from the CHIANTI
atomic database, version 5 \citep{dere97, landi06}. These data include
levels with principal quantum number $n \leq 5$ for the H-like ions and
for states up to the 1s5g configuration for He-like ions.
Ni {\sc xxv} and {\sc xxvi} are included only for the calculation of
the ionization
balance; currently no bound-bound transitions for these ions are treated.

For Fe, atomic data were taken from
{TIPBASE\footnote{http://vizier.u-strasbg.fr/tipbase/home.html}} 
which provides atomic data from the {IRON} project for ions of astrophysical interest.
These data also contain levels for  $n \leq 5$ of Fe~{\sc xxvi} but
include more highly excited configurations of Fe~{\sc xxv} than
CHIANTI (up to the 1s10h configuration). Atomic data for Fe~{\sc
xxiii} and {\sc xxiv} were also taken from TIPBASE; however, no states
involving excitation of electrons from the 1s shell are included for
either of these ions.

Ground configuration photoionization cross-sections were described by
the fits from \citet{verner96}, except for Ni for which fits from
\citet{verner95} were adopted. For excited states, photoionization
cross-sections were computed using a hydrogenic approximation for all
ions. Since bound-free absorption from excited states was not included
in the MC simulations, these excited state
cross-sections are only needed for the computation of macro atom
transition probabilities (see Section~\ref{sect_reemiss}).
The fraction of recombinations which go directly to the ground state,
$\zeta$, was obtained from the hydrogenic calculations of \citet{martin88}.

\section{Example calculation}
\label{sect_example}

In this section we present detailed results pertaining to one
particular instance of our model.

\subsection{Model parameters}

The complete set of adopted input
parameters for our example model is give in Table~\ref{tab_param}.

The black-hole mass, source X-ray
luminosity and primary power-law index 
are all chosen to be reasonable for Mrk~766, the object for which we
make a detailed comparison in Section~\ref{sect_mrk766} 
($M_{\mbox{\scriptsize bh}} \sim 4.3 \times 10^6$~M$_{\odot}$ [\citealt{wang01}]; 
$L_{X} \sim 10^{43}$~ergs~s$^{-1}$ [\citealt{pounds03b}]; 
$\Gamma \sim 2.38$ [\citealt{miller07}]).
We assume that $f_{v} = 1$, which is characteristic of radiatively
driven flows. Furthermore, since we wish to consider outflow features
with shifts corresponding to $v_{\infty} \simgt 0.1 c$, we require
flow launching radii 
$\simlt 200 r_g$ -- hence we adopt $r_{\mbox{\scriptsize min}} = 100 r_g$ and $r_{\mbox{\scriptsize max}}
= 150 r_g$.

Physically, $R_{v}$ and $\beta$ should depend on the acceleration
mechanism and location -- however, since such properties are unknown,
we adopt $R_{v} = r_{\mbox{\scriptsize min}}$ and $\beta = 1$,
as appropriate for the simple case of an acceleration which occurs on the scale
of the system. It is assumed that $v_{0} \ll v_{\infty}$ such that
$v_{0} = 0$ can be adopted in the numerical simulations.

For the example model, we adopt a polar opening angle of 45~deg
(i.e. $d = r_{\mbox{\scriptsize min}}$), a moderate mass-loss rate ($\dot{M} = 0.1$ M$_{\odot}$
yr$^{-1}$) and electron temperature $T_e = 3 \times 10^6$~K. The
effects of varying these parameters will be discussed in Section~\ref{sect_grid}.

\begin{table}
\caption{Inputs parameters for the example model. The upper portion
  gives the adopted physical parameters and the lower portion gives
  numerical parameters.}
\label{tab_param}
\begin{tabular}{ll}\\ \hline
Parameter & Value \\ \hline \hline 
rad. source luminosity (2 -- 10 keV), $L_{X}$ & $10^{43}$ ergs~s$^{-1}$\\
source power-law photon index, $\Gamma$ & $2.38$ \\
mass of central object, $M_{\mbox{\scriptsize bh}}$ & $4.3 \times 10^6$ M$_{\odot}$ \\
inner launch radius, $r_{\mbox{\scriptsize min}}$ & 100 r$_g = 6.4 \times 10^{13}$ cm \\
outer launch radius, $r_{\mbox{\scriptsize max}}$ & $1.5 r_{\mbox{\scriptsize min}}$ \\
distance to wind focus, $d$ & $r_{\mbox{\scriptsize min}}$ \\
terminal velocity parameter, $f_{v}$ & 1.0 \\
velocity scale length, $R_{v}$& $r_{\mbox{\scriptsize min}}$\\
velocity exponent, $\beta$& 1.0 \\
launch velocity, $v_{0}$ & 0.0 \\
wind mass-loss rate, $\dot{M}$ & 0.1 M$_{\odot}$
yr$^{-1}$ \\
mass-loss exponent, $k$ & -1.0 \\
electron temperature, $T_{e}$ & $3 \times 10^6$~K \\ \hline \hline
extent of grid & $2 \times 10^{15}$ cm \\
3D Cartesian RT grid size & $100 \times 100 \times 100$ \\
2D wind grid & $100 \times 100$ \\ 
Number of MC quanta & $2.4 \times 10^8$ \\ \hline
\end{tabular}
\end{table}

\subsection{Computed ionization state}

\begin{figure*}
\epsfig{file=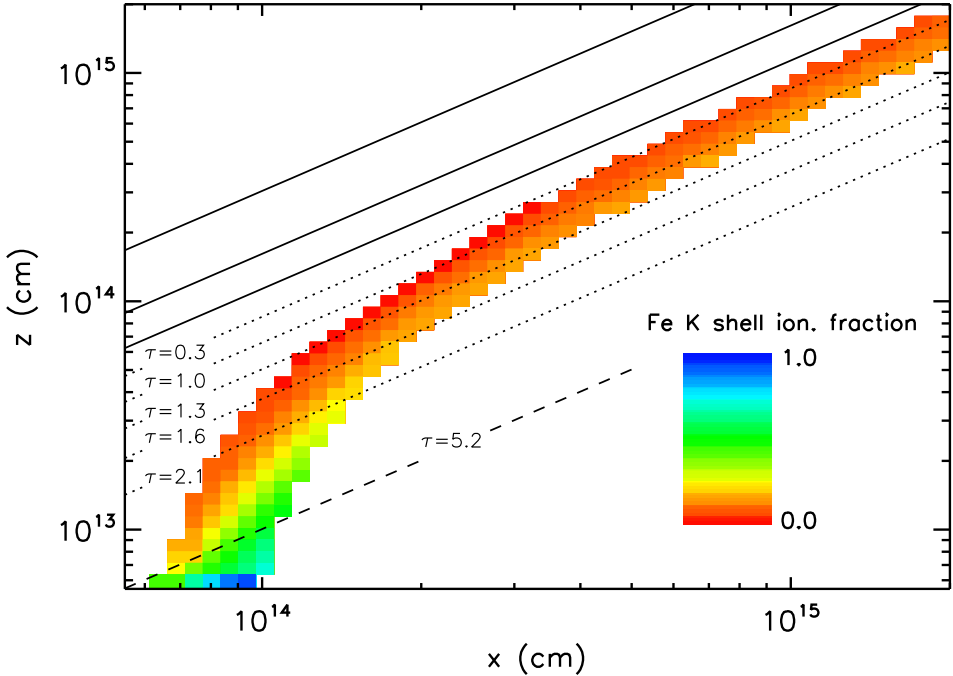, width=14cm}
\caption{
The computed ionization fraction for K-shell Fe (i.e. Fe~{\sc xxv} +
{\sc xxvi}). Only the positive $xz$-plane is shown. Note the
logarithmic axes and the difference in scale factor on the $x$- and
$z$-axes. 
{ 
The blue-red colours indicate the ion fraction on a
linear scale}
-- red shades represent regions where the wind is
almost fully ionized (i.e. Fe~{\sc xxvii} is dominant).
The black lines indicate lines-of-sight from the X-ray source (the
coordinate origin); the lines-of-sight drawn are those at the
mid-points of the nine angular bins used for the spectra shown in
Fig.~\ref{fig_spec}. The solid lines are lines-of-sight that never
intersect the outflow (corresponding to the top three panels in Fig.~\ref{fig_spec}); the dotted lines intersect the outflow and yield 
spectra with narrow absorption features (fourth to eighth panels in Fig.~\ref{fig_spec}); the dashed line intersects
the outflow and produces a spectrum without narrow absorption features
(last panel in Fig.~\ref{fig_spec}). 
{
The lines-of-sight which intersect the
flow are labelled with their integrated Compton optical depth ($\tau$)
from the continuum source (coordinate origin) to the observer.
}
}
\label{fig_ion}
\end{figure*}

The ionization state is very important since it
determines the distribution of line opacity within the
outflow. Fig.~\ref{fig_ion} shows the computed ionization fraction of
Fe K-shell ions (i.e. Fe~{\sc xxv} + {\sc xxvi}) in the example model.

There is significant variation in the ionization state in the
wind. The edge of outflow closest to the rotation ($z$-) axis is
most highly ionized since it sees unattenuated X-ray
radiation from the continuum source. 
In the outermost parts of the wind the ionization state is high and
relatively uniform: although far from the X-ray source,
these regions have low density which disfavours recombination.
Across the wind, there is a significant 
ionization gradient since
each layer progressively shields those below it from the X-ray
source. The very lowest ionization material occurs near the $xy$-plane 
on the outer edge of the wind -- this region has both the
highest densities and the most effective shielding from the X-ray
source.

\subsection{Computed spectra}

\begin{figure*}
\epsfig{file=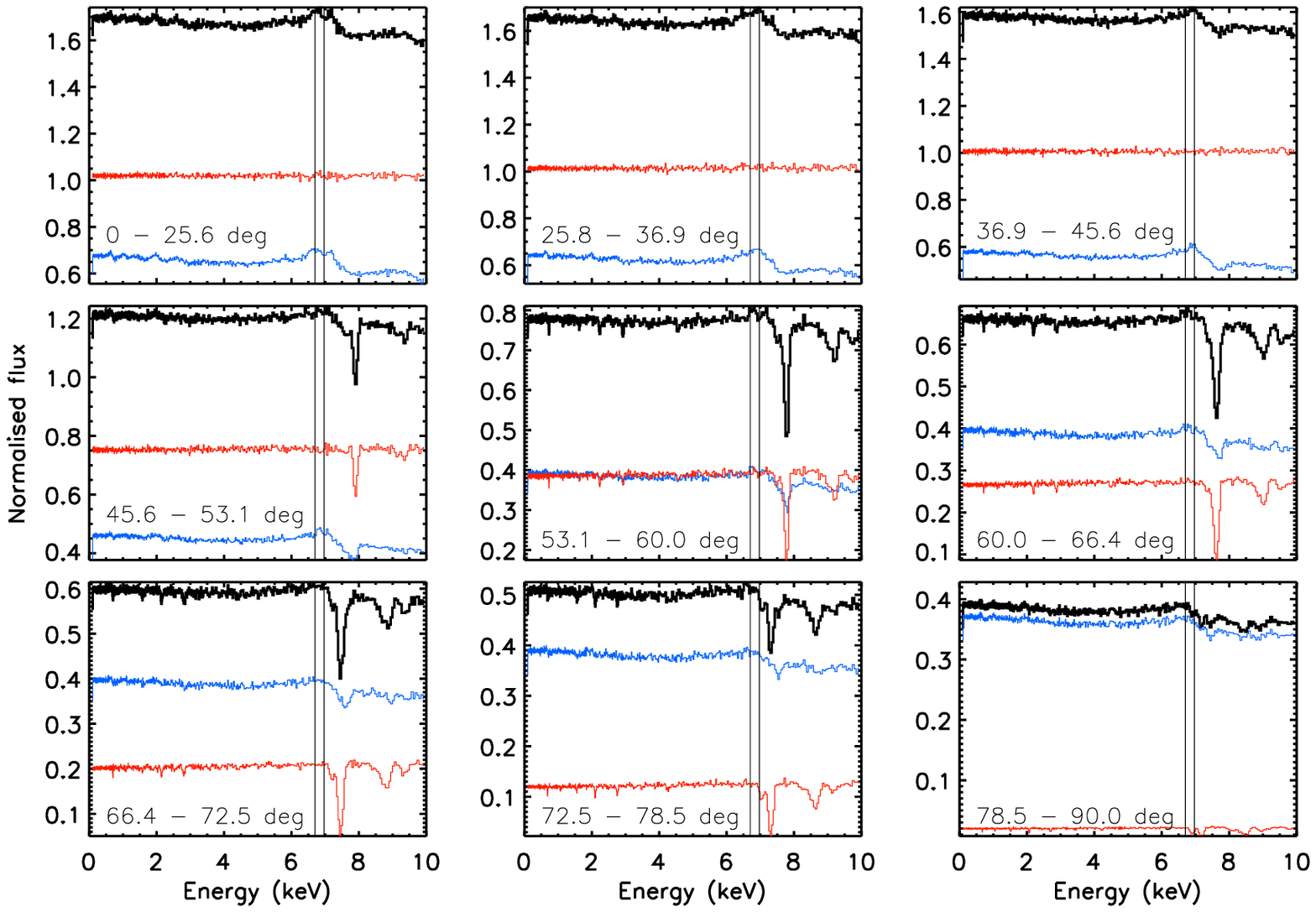, width=15cm}
\caption{
Spectra computed for the example model for different lines-of-sight.
Each panel shows the spectrum in a different inclination angle bin
(the angular bin sized are indicated in the figure, measured relative
to the polar axis). Note that the ordinate range is different in each
panel and does not extend to zero.
The total spectrum is plotted in black. The
spectrum is separated into direct and scattered components and these
are overplotted in red and blue, respectively. 
All the spectra are shown normalised to the input primary power-law spectrum.
The solid vertical
lines indicate the rest energies of Fe~K$\alpha$ for the {\sc xxv} and
{\sc xxvi} ionization stages.}
\label{fig_spec}
\end{figure*}

Fig.~\ref{fig_spec} shows spectra computed for the example
model. These were obtained by binning the emergent MC quanta by angle
relative to the polar axis ($\theta$). The first eight angular bins shown each
cover equal solid angle, specifically they encompass $\Delta \cos
\theta = 0.1$ while the ninth covers the equatorial range $0 < \cos
\theta < 0.2$.

During the MC simulations, the number of interactions was recorded for
each quantum. This information has been used to divide the spectra into
``direct'' (meaning quanta which underwent no interactions) and
``scattered'' components (note that this includes packets that
underwent {\it any} number of {\it any} type of physical event -- it
does not only include true scattering events). These component spectra
are also plotted in the figure.

It is apparent that the outflow in the example model
affects the spectrum as observed from any
line-of-sight. The typical apparent luminosity varies by more than a 
factor of four -- the flux level is much higher close to the polar
axis than near the equatorial plane. 
This is a consequence of scattering in the flow, primarily by free
electrons. For small angles relative to the polar axis 
the source radiation escapes directly, unhindered by the wind. At high
inclination angles, the photons enter the wind and most are
scattered, thereby changing their
direction of propagation. Compared to the naked source, 
this enhances the radiation for the polar direction and
suppresses it close to the equator.
Also, as one would expect, the scattered
radiation field is softer than the direct such that the combined
spectrum is also slightly softer.

For viewing angles less than
45~deg 
the spectrum is unobscured by the outflow and is supplemented by
scattered radiation which, for this model, can have more than 50 per
cent as high a
flux as the direct component. 
These lines-of-sight are indicated by the solid lines in
Fig.~\ref{fig_ion} and the corresponding spectra are shown in the 
top three panels of Fig.~\ref{fig_spec}.
The scattered radiation field shows
signatures of scattering in spectral lines -- a 
P-Cygni-type line profiles is apparent in the Fe~K$\alpha$ line and there are
weaker but similar features arising from the K$\alpha$ lines of lighter 
elements at softer energies.

At intermediate viewing angles (dotted lines-of-sight 
in Fig.~\ref{fig_ion}), the line-of-sight intersects the flow.
{The
Compton optical depth along these sight lines ($\tau$) is
of order unity (see labels in Fig.~{\ref{fig_ion}}) meaning that a 
significant fraction of the continuum radiation 
emitted in these directions is scattered by the flow.
}
Discreet narrow blueshifted 
absorption lines appear in these
spectra (fourth to eighth panels of Fig.~\ref{fig_spec}). 
The blue-shift which manifests in any
particular spectrum depends 
on the component of outflow velocity which lies along the
line-of-sight
and hence
on the inclination.
In our model, the blueshift increases as $\theta$ decreases.
The scattered component also contains 
absorption features, albeit weaker and broader than those in the
direct spectrum. In addition, it includes
emission which results from line-scattering throughout the
model and is thus significantly velocity-broadened.

When viewed from high inclination angle (dashed line-of-sight
in Fig.~{\ref{fig_ion}} and last panel of Fig.~{\ref{fig_spec}}), 
the flow is significantly optically thick and the spectrum is totally
dominated by scattered radiation.
These spectra show weak emission lines and weak,
relatively broad absorption lines.

Thus our example model demonstrates that the simple disk wind
geometry adopted can naturally produce narrow, blueshifted absorption
features, as required. However, it also makes clear that such features
are not necessarily produced and that depending on the inclination
angle a variety of other features, including relatively broad emission
or absorption, may be present. 

In the example model, the greatest 
contrast of the line features to the continuum, 
is only about 40 per cent.
There are lines-of-sight (e.g. $\theta \sim$ 75~deg) in 
which the direct component shows narrow absorption features which are
close to black at line centre.
However, in the total spectrum the contrast is always reduced by
addition of the
scattered component in which all features are smeared by
velocity broadening. 
We note that 
our assumption of a stationary, 
smooth flow is likely to maximise this effect since it will generally
overestimate the relative strength of scattered to direct radiation
for lines-of-sight in which absorption features occur.

\section{Model grid}
\label{sect_grid}

Having discussed the complete spectra from the example model
(Section~\ref{sect_example}), we now explore the effects of some
critical model parameters on the computed spectra. 
To do this we have constructed a grid of 45 models and computed
spectra for each in 10 viewing-angle bins. The parameters which were
varied between models, and their respective effects on the spectrum,
are discussed in the sections below.

In order to elucidate our discussion,
we have performed simple measurements of 
the properties of the Fe~K absorption features in the spectra. 
First, for each spectrum a continuum spectrum was obtained by masking
out the strong K$\alpha$ absorption lines and then median filtered to 
suppress MC noise. The equivalent width (EW) of any Fe~K$\alpha$ absorption
was then measured by numerical integration of the normalised spectrum.
The spectra with absorption EWs $> 40$eV were then classified
by eye into one of three categories:

\begin{enumerate}

\item spectra with a single, clear Fe K$\alpha$ absorption line;

\item those having two, well-separated Fe K$\alpha$ absorption lines
  (one due to Fe~{\sc xxv} and the other due to Fe~{\sc xxvi});

\item those with K$\alpha$ absorption in both  Fe~{\sc xxv} and
  Fe~{\sc xxvi}
  but in which the two lines are strongly blended.

\end{enumerate}
For the lines in categories (i) and (ii), the blueshift line velocity
was found by measuring the photon energy of deepest
absorption and an approximate
full-width-at-half-maximum (FWHM) was determined for the absorption
feature by fitting a Gaussian profile. 
For the spectra in class (iii), Fe~K$\alpha$
velocities and FWHM were not extracted. Although these line
properties do not adequately describe all properties of the spectra,
they provide a convenient quantification of the spectra for the
discussions in the sections below.

\subsection{Mass-loss rate, $\dot{M}$}
\label{sect_mdot}

\begin{figure}
\epsfig{file=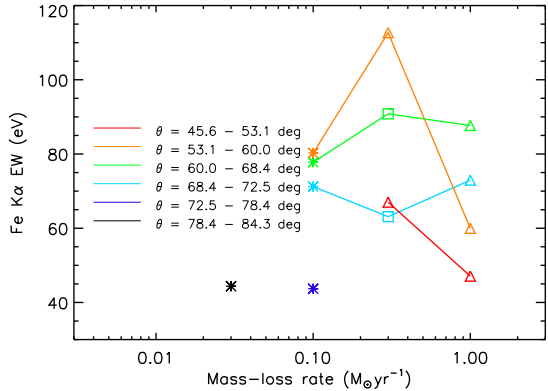, width=8cm}
\caption{
Fe~K$\alpha$ absorption EWs for models with different $\dot{M}$-values (all
other parameters held fixed). The star, triangle and box symbols
indicate spectra in which there is a single, clear Fe~K$\alpha$
absorption line, those in which there is a pair of separated
Fe~K$\alpha$ lines and those in which the Fe~{\sc xxvi} and {\sc xxv}
components are strongly blended, respectively.
The colours identify spectra based on the inclination of the observer
line-of-sight to the polar axis ($\theta$). Note that only spectra
with Fe~K$\alpha$ absorption EWs $> 40$eV are included.}
\label{fig_mdot}
\end{figure}

The wind mass-loss rate is the most important model parameter which
one would wish to constrain observationally. To demonstrate its
effect, we have computed spectra for models with five $\dot{M}$ values
(0.01, 0.03, 0.1, 0.3 and 
1.0 M$_{\odot}$~yr$^{-1}$); the Fe~K$\alpha$ absorption EWs for these
models are shown in Fig.~\ref{fig_mdot}. 
{ 
For Mrk~766, this range of mass-loss rates explores the region around 
the mass
accretion rate estimated from the bolometric luminosity
$\dot{M}_{\mbox{\scriptsize acc}} \approx {L_{\mbox{\scriptsize
bol}}}/{0.054 c^2} \sim 0.1$~M$_{\odot}$~yr$^{-1}$ 
adopting the bolometric
luminosity ($L_{\mbox{\scriptsize bol}} = 3 \times
10^{44}$~ergs~s$^{-1}$) from \citet{hao05} and the 
radiative efficiency for a Schwarzschild black hole \citep{shapiro83}.
}

$\dot{M}$ controls the density in the wind and therefore
affects both the ionization state and all line or continuum optical
depths. At the low end of our parameter range, the ionization state is
high enough that absorption by Fe~{\sc xxvi} is much more important
than Fe~{\sc xxv}. Increasing $\dot{M}$
leads to an increase in absorption EWs, as one would expect from
the increase in opacity. Also for small $\dot{M}$, strong 
absorption is only present at large $\theta$ since only the very base
of the flow is dense enough to present significant opacity.

At intermediate $\dot{M}$, absorption appears across a wider range of
$\theta$ and becomes generally stronger. 
Increased densities reduce the degree of ionization such that 
absorption by both Fe~{\sc xxv} and Fe~{\sc xxvi} K$\alpha$
becomes more common.
Absorption at very high
inclination angles becomes less significant since at these angles, the
spectra are increasingly dominated by scattered radiation which
washes out any narrow absorption features imprinted on the direct
radiation.

For the highest mass-loss rates considered, all the spectra contain
absorption by both Fe~{\sc xxvi} and Fe~{\sc xxv} K$\alpha$ 
and scattered emission is very important at all
inclination angles. This means that the EWs are mostly lower in the
$\dot{M} = 1$ than $\dot{M} = 0.3$~M$_{\odot}$~yr$^{-1}$ spectra.

\subsection{Wind geometry, $d$}
\label{sect_d}

Within our simple wind prescription, $d$ controls the opening angles of
the outflow and therefore determines both the angular range of
lines-of-sight which pass through the wind and the relative
orientations of the outflow and rotational velocity fields. To explore
the effect of geometry on the spectrum, we have extended our grid of
models to include $d$-values of 0.33~$r_{\mbox{\scriptsize min}}$ and 3~$r_{\mbox{\scriptsize min}}$,
thereby producing geometries which are, respectively, more equatorial
and more polar than the example model. For these $d$-values, we have
computed models with the same set of $\dot{M}$-values used in
Section~\ref{sect_mdot}, keeping all other parameters fixed to those
of the example model.

\begin{figure}
\epsfig{file=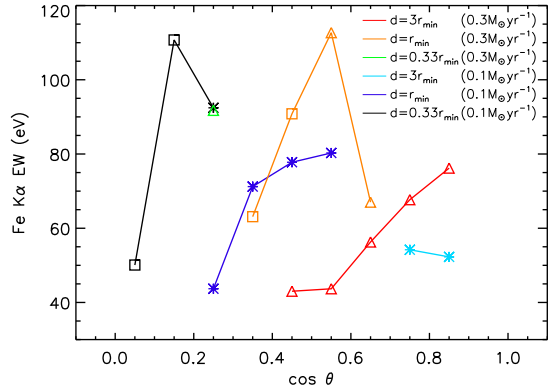, width=8cm}
\caption{
Fe~K$\alpha$ absorption EWs versus line-of-sight inclination for
models with different $d$- and $\dot{M}$-values (all
other parameters held fixed). The plotting symbols are as described
for Fig.~\ref{fig_mdot}. Note that only spectra
with Fe~K$\alpha$ absorption EWs $> 40$eV are included.}
\label{fig_d}
\end{figure}

Fe~K$\alpha$ absorption EWs are plotted for a sample of these models
in Fig.~\ref{fig_d}. This shows the
role of the wind opening angles in determining which lines-of-sight
intersect the outflow in regions where optical depths are high enough
to cause line absorption.

\begin{figure}
\epsfig{file=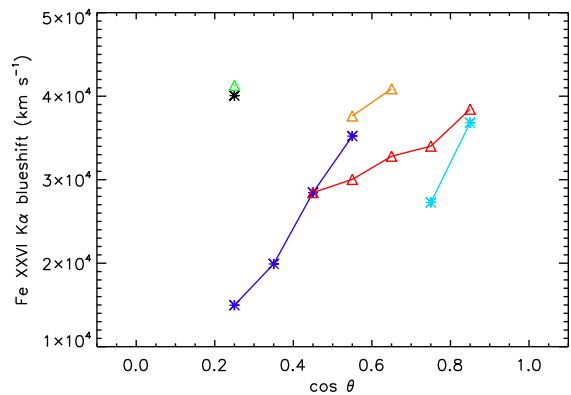, width=8cm}
\caption{
The blueshift of the 
Fe {\sc xxvi}~K$\alpha$ line
versus line-of-sight inclination for
models with different $d$- and $\dot{M}$-values.
The plotting symbols and colors are as described for
Fig~\ref{fig_d}. Note that only spectra
with Fe~K$\alpha$ absorption EWs $> 40$eV are included.}
\label{fig_d_vel}
\end{figure}

The wind opening angles and line-of-sight orientation also affect 
the observed Fe~K$\alpha$ blueshift.
Fig.~\ref{fig_d_vel} shows the Fe~{\sc xxvi} blueshift velocity for
the same models used in Fig.~\ref{fig_d}, 
excluding spectra with
blended Fe~{\sc xxv} and {\sc xxvi} K$\alpha$ absorption.
As mentioned in Section~\ref{sect_example},
the 
variation in blueshift
arises from line-of-sight projection of the outflow velocity 
and the effect is large: there can be more than a factor of two
spread in the apparent blueshift.
Since the maximum outflow velocity is

\begin{equation}
v_{\infty} = \sqrt{ \frac{2 G M_{\mbox{\scriptsize bh}}}{r_{\mbox{\scriptsize min}}}} = 0.14 c = 4.2 \times
10^4 \mbox{km s$^{-1}$} 
\end{equation}
for all the models discussed here, the variation in blueshift is
entirely a consequence of orientation.
Thus, while the apparent blueshift can
be as large as the terminal outflow velocity, it is often
significantly smaller.

\subsection{Inner launch radius, $r_{\mbox{\scriptsize min}}$}
\label{sect_rm}

\begin{figure}
\epsfig{file=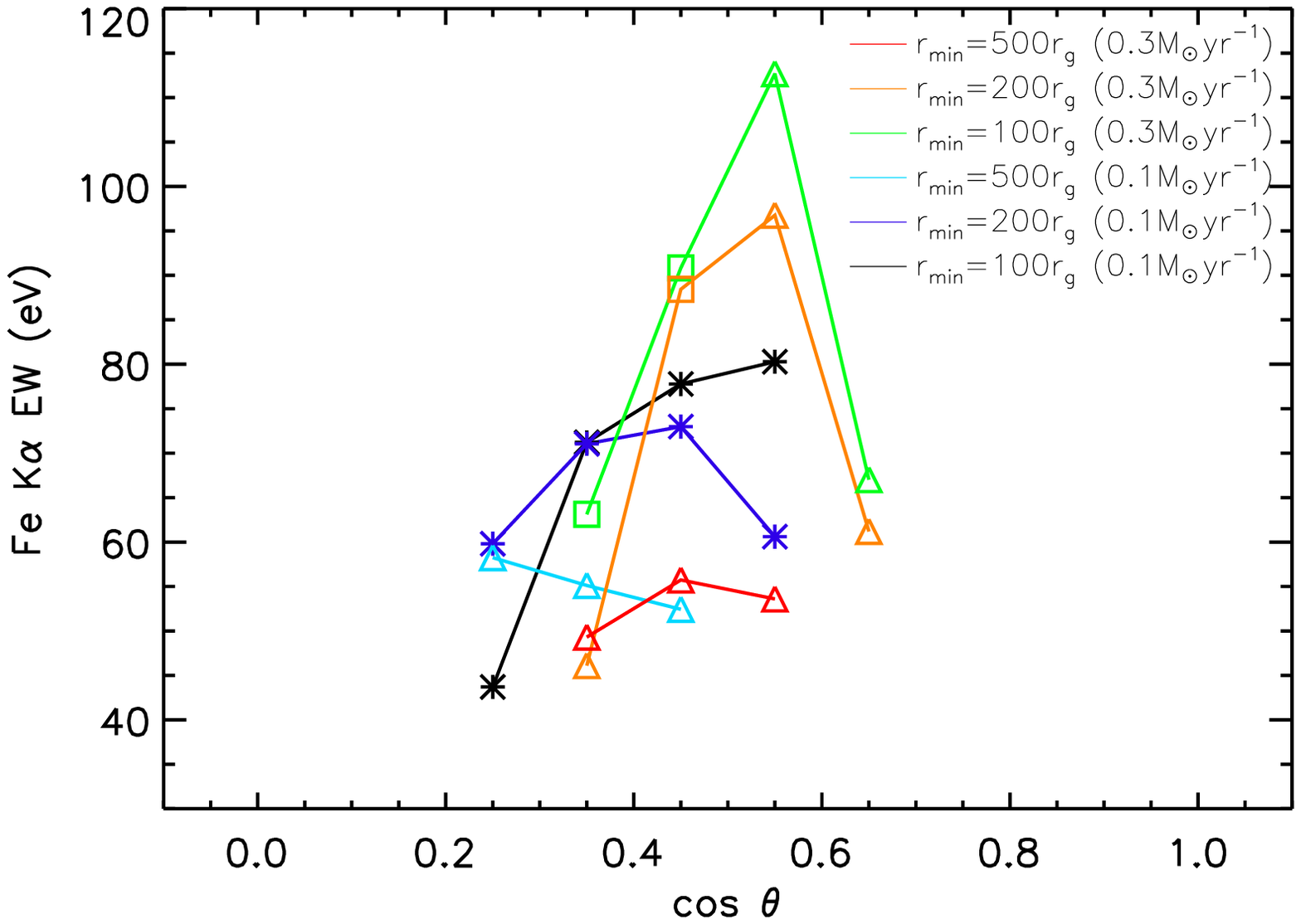, width=8cm}
\caption{
Fe~K$\alpha$ absorption EWs versus line-of-sight inclination for
models with different $r_{\mbox{\scriptsize min}}$- and $\dot{M}$-value. The plotting symbols are as described
for Fig.~\ref{fig_mdot}. Note that only spectra
with Fe~K$\alpha$ absorption EWs $> 40$eV are included.}
\label{fig_rm}
\end{figure}

The inner launch radius of the wind,  $r_{\mbox{\scriptsize min}}$, determines how far
the wind material is from the X-ray source and
affects the wind density since the surface area of
annuli in the disk plane increases with $r$. To investigate its
effect, we have computed spectra for models with $r_{\mbox{\scriptsize min}}$ larger than
in the standard model ($r_{\mbox{\scriptsize min}} = 200 r_{g}$ and $500
r_{g}$) adopting the same range of $\dot{M}$-values used in
Section~\ref{sect_mdot}. For all these models, we also varied other
parameters as necessary to preserve the
relationships used for the standard model (i.e. $r_{\mbox{\scriptsize max}} = 1.5
r_{\mbox{\scriptsize min}}$; $d = r_{\mbox{\scriptsize min}}$; $R_{v} = r_{\mbox{\scriptsize min}}$, $f_{v} = 1$).
Fig.~\ref{fig_rm} shows Fe~K$\alpha$ absorption EWs computed as a
function of viewing angle for six of these models.

Changing $r_{\mbox{\scriptsize min}}$ affects the typical range of viewing angle for
which absorption features are strong, favouring smaller $\cos \theta$ for 
larger $r_{\mbox{\scriptsize min}}$. For
fixed $\dot{M}$, increasing $r_{\mbox{\scriptsize min}}$ tends to make the absorption
lines weaker, a consequence of the reduced density in the flow.

\subsection{Electron temperature, $T_{e}$}

\begin{figure}
\epsfig{file=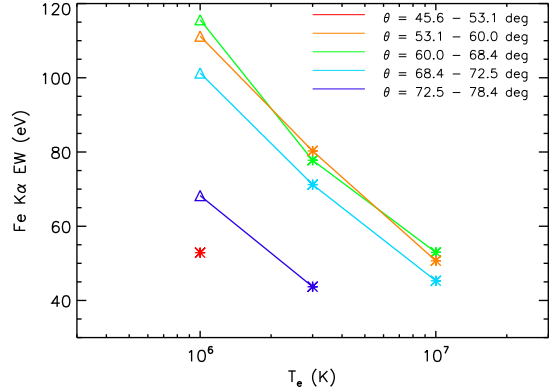, width=8cm}
\caption{
Fe~K$\alpha$ absorption EWs for models with different $T_{e}$-values (all
other parameters held fixed). The plotting symbols are as described
for Fig~\ref{fig_mdot}. Note that only spectra
with Fe~K$\alpha$ absorption EWs $> 40$eV are included.}
\label{fig_te}
\end{figure}

Under our assumptions, the only important role played by the electron
temperature is in determining the ionization fractions. To illustrate
its effect, we have used models which span the same $\dot{M}$-grid as
discussed in the section above but which adopt $T_{e}$ both lower ($10^6$~K)
and higher ($10^7$~K) than used in the example model. EWs for a
subset of these models, namely all those having 
$\dot{M} = 0.1$~M$_{\odot}$~yr$^{-1}$ are shown in Fig.~\ref{fig_te}.

The dependence on $T_{e}$ is these models is very simple. As $T_{e}$ increases, the
wind becomes more ionized, reducing the H- and He-like ion populations
in favour of the fully ionized state. This reduces the line opacities
and causes the absorption EWs to decrease. Absorption in both Fe~{\sc
xxv} and {\sc xxvi} is most common at low $T_{e}$ since the Fe~{\sc
xxv} population is depleted more rapidly than that of 
Fe~{\sc xxvi} as the wind becomes more ionized.

\subsection{Primary power-law index, $\Gamma$}

\begin{figure}
\epsfig{file=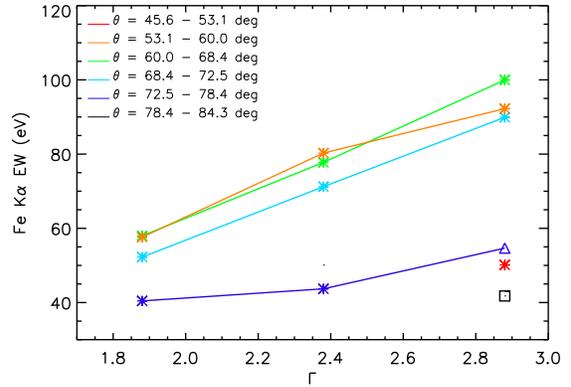, width=8cm}
\caption{
Fe~K$\alpha$ absorption EWs for models with different $\Gamma$-values (all
other parameters held fixed). The plotting symbols are as described
for Fig~\ref{fig_mdot}. Note that only spectra
with Fe~K$\alpha$ absorption EWs $> 40$eV are included.}
\label{fig_G}
\end{figure}

The source power-law index determines the fraction of the X-ray flux
which is hard enough to contribute to the ionization of Fe~{\sc xxv}
and {\sc xxvi} and therefore has an effect on the Fe~K$\alpha$
absorption EWs which is qualitatively
similar to that of $T_{e}$. In Fig.~\ref{fig_G} we
show Fe~K$\alpha$ EWs for models with $\dot{M} =
0.1$~M$_{\odot}$~yr$^{-1}$ adopting $\Gamma = 1.88, 2.38$ and
2.88, with other parameters as in the
example model. As expected, higher $\Gamma$s lead to larger EWs since 
there are fewer Fe K-shell ionizing photons and hence the
ionization state is lower.

Note that the total 2 -- 10 keV luminosity ($L_{X}$) is constant here 
-- although we have not explicitly explored the effect of changing 
$L_{X}$, it is clear that it will have a similarly simple effect on the
ionization state.

\section{Observable properties}
\label{sect_obs}

In Section~\ref{sect_grid} we constructed a grid of models to explore
the effects of various wind properties on the Fe~K$\alpha$ absorption
line properties. However, since the relevant wind parameters are
not known a priori it is instructive to investigate the properties of
the models in the space defined by observable quantities.
Given the typical data quality and limited spectral resolution of
current X-ray telescopes, the primary observables pertaining to
K$\alpha$ absorption lines are the EW and the velocity shift of the
lines.

\begin{figure}
\epsfig{file=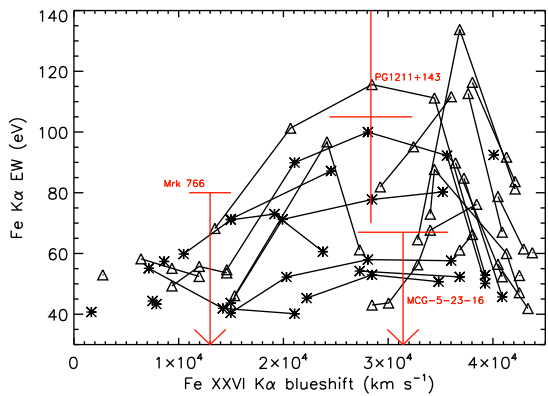, width=8cm}
\caption{
Fe~K$\alpha$ absorption EWs versus the Fe~{\sc xxvi} blueshift.
All spectra computed from our grid of models (450 spectra in total)
with unblended Fe~K$\alpha$ absorption with EW $> 40$~eV
are included
in the plot. The star symbols represent model spectra in which a
single Fe~K$\alpha$ absorption line is dominant. Triangles indicate
spectra with two, unblended  Fe~K$\alpha$ absorption lines.
The solid lines connect points for the same model but viewed from
different inclination angular bins.
The red cross indicates the Fe K$\alpha$ absorption line 
reported by \citet{pounds03} for PG1211+143 while the red bars
indicate the range of EWs obtained with the time sliced observations
of Mrk~766 \citep{turner07} and MCG-5-23-16 \citep{braito07}.
}
\label{fig_ewvsshift}
\end{figure}

Of the 450 spectra which were generated with our grid of models 
(Section~\ref{sect_grid}), 101 were classified as having Fe~K$\alpha$
absorption EWs of $> 40$ eV and, of these, 79 were classified as having
either a single or an unblended pair of absorption features such that
a meaningful velocity blueshift could be extracted from the spectrum.
Fig.~\ref{fig_ewvsshift} shows the distribution of Fe
K$\alpha$ absorption EWs versus the Fe~{\sc xxvi} blueshift for these
79 spectra.

There are a
wide variety of K$\alpha$ absorption features in our models. 
EWs of more than
100~eV are rare but do occur; the maximum EW produced by any of the models is
$\sim 135$~eV. As one would expect, the EW tends to be larger when
two lines are present although there are cases where 
a single line achieves an EW of nearly 100~eV. As already
highlighted in Section~\ref{sect_d}, a very wide range of
blueshift velocity is predicted. 
Typically, the maximum EWs is larger
when the blueshift is larger; but there is significant variation
in EW at all blueshifts.
 In general, there are many sets of wind
 parameters that can produce similar EWs and blueshifts such that
 K$\alpha$ properties cannot uniquely discriminate 
 between our models. Other potential outflow
 signatures are discussed in Section~\ref{sect_beyondka}.

As a simple comparison, points representing reported detections of
narrow blueshifted K$\alpha$ absorption in the spectra of three real AGN
(PG1211+143, Mrk~766 and MCG-5-23-16; see below)
are overplotted in Fig.~\ref{fig_ewvsshift}. Observations clearly
show that outflow absorption features are significantly time
variable (e.g. \citealt{risaliti05, miller07, turner07, braito07})
and thus one must be cautious when comparing them directly to the
results of our steady-state models -- the development of 
multi-dimensional 
models involving clumpy or structured flows is necessary for a
detailed comparison.

\citet{pounds03} first
reported blueshifted K$\alpha$ absorption from H- or He-like Fe in the
spectrum of PG1211+143 (see also \citealt{pounds05,pounds06,pounds07}
for further discussion of this interpretation of the PG1211+143
spectrum); they report an absorption EW of $105 \pm 35$~eV
\citep{pounds05} and a blueshift of either 0.08 -- 0.10 $c$ 
\citep{pounds05} or 0.13 - 0.15 $c$ \citep{pounds06}, depending on
whether one assumes the observed features is owing to Fe {\sc xxvi}
or {\sc xxv}. The point plotted in Fig.~\ref{fig_ewvsshift} for this
object assumes the Fe {\sc xxvi} interpretation and the measurement
uncertainties quoted by \citet{pounds05}.

Two distinct
absorption features (6.9 and 7.2~keV) were reported in 
spectra of Mrk~766 \citep{turner07, miller07}, at 6.9 and 7.2~keV. 
Although the nature and relationship between these
features is unclear, one possible interpretation is that they
correspond to blueshifted
absorption by Fe~{\sc xxv} and Fe~{\sc xxvi}, respectively -- in this
case, the blueshift velocity of both lines is $\sim
13,000$~km~s$^{-1}$ \citep{miller07}. 
\citet{turner07} showed that the strength of these features varies
significantly in time.
The largest fluxes of the absorption features
occurred in their time slice 9 (see their fig. 2) and had an 
EW $\sim 80$~eV each. In some time slices, the line fluxes are
consistent with zero.
In Fig.~\ref{fig_ewvsshift}, the range of Mrk~766
absorption feature properties is marked adopting the
13,000~km~s$^{-1}$-outflow interpretation (i.e. identifying the
7.2~keV feature with Fe~{\sc xxvi}) and the EW range of 0 -- 80~eV as
motivated by the time-sliced data.

\citet{braito07} detected a variable absorption feature at 7.7~keV in
the spectrum of MCG-5-23-16. 
This feature is also indicated in Fig.~\ref{fig_ewvsshift}, adopting
the upper bound of on its EW range based on
their analysis of the one of five 20~ks time
intervals in which
the line was most clearly detected (see their table 4).
The velocity shift was computed
assuming that Fe~{\sc xxvi} is dominant.

In terms of shift and
range of EW, the absorption features
in PG1211+143, Mrk~766 and MCG-5-23-16 
lie close to the space explored by our grid
 of models (Fig.~\ref{fig_ewvsshift}). 
Thus wind models such as those we have investigated are
 promising starting points for explaining the K$\alpha$ absorption in these
 observations.

Blueshifted K$\alpha$ absorption features have also been reported in
NGC~1365 \citep{risaliti05}. In this case, the shifts are
considerably smaller, $\sim$5000~km~s$^{-1}$ at most. Although our
grid of models does include some spectra with features at these low
velocities (see Fig.~\ref{fig_ewvsshift}), the EWs in the computed
spectra are below those observed by a
significant factor (the combined Fe K$\alpha$ EWs reported by
\citealt{risaliti05} are in the range 200 -- 300~eV, lying off the
scale of Fig.~\ref{fig_ewvsshift}). 
Even lower-blueshift 
highly-ionized absorption has been
reported in MCG-6-30-15 \citep{young05,miniutti07,miller08} and 
NGC~3516 \citep{turner08}.
The simplest
way in which such strong, relatively low-velocity 
features might be produced with our model would be by considering
lower terminal flow velocities but it may also be possible to obtain
deeper absorption by considering flows which are significantly
clumped. Detailed study of these possibilities lies beyond the scope
of this paper, but we do note that even the NGC~1365 features are
within a factor of $\sim 5$ in EW of those obtained with our grid of models.

\begin{figure}
\epsfig{file=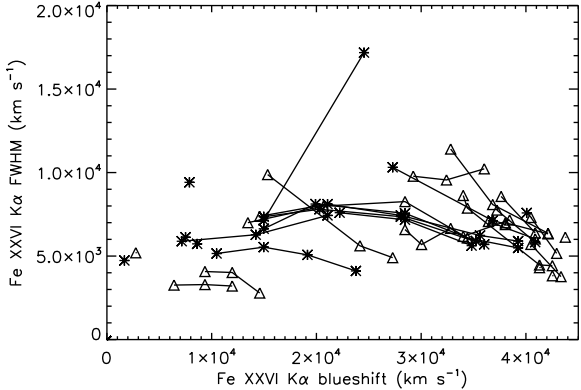, width=8cm}
\caption{
Fe~K$\alpha$ FWHM versus the Fe~{\sc xxvi} blueshift.
The plot symbols are the same as in Fig.~\ref{fig_ewvsshift}.
}
\label{fig_widvsshift}
\end{figure}

In principle, the model spectra indicate that the absorption line
shapes are complex and therefore provide information on the outflow
properties. However,
given the limited sensitivity and spectral resolution of current X-ray
telescopes, determining line shapes is virtually impossible -- at best,
only a rough limit on the characteristic linewidth is available.
For comparison with such limits, Fig.~\ref{fig_widvsshift} shows the
FWHM determined from our spectra (by Gaussian fitting) versus the
Fe~{\sc xxvi} blueshift for the same models as plotted in 
Fig.~\ref{fig_ewvsshift}.

The FWHM varies between about $3.5 \times 10^{3}$ and $1.8 \times
10^4$~km~s$^{-1}$, with values around 5 -- 7 $\times 10^3$~km~s$^{-1}$
being most common.
For all our model spectra in which the blueshift is greater than 
10$^4$~km~s$^{-1}$, the FWHM is smaller than the blueshift meaning that
the absorption does not
extend down to the rest energy of the line. As one
would expect, the
largest linewidth ($\sim 1.8 \times 10^4$~km~s$^{-1}$) occurs in
model where $d$ has the smallest value considered
($d = 0.33$) and for a viewing direction which has a large pathlength
through the flow.

To our knowledge,
there are currently no firm measurements of linewidths for narrow blueshifted 
Fe~K$\alpha$ absorption features in AGN spectra. For PG1211+143,
\cite{pounds03} placed a limit of $< 1.2 \times 10^4$~km~s$^{-1}$ on
the Fe K$\alpha$ linewidth, which is consistent with the majority of
the measurements from our spectra shown in Fig.~\ref{fig_widvsshift}.
\cite{braito07} found a best-fit line width of $\sigma = 200 \pm
100$~eV (FWHM $= 1.8 \times 10^{4}$~km~s$^{-1}$) for the absorption
feature in MCG-5-23-16. However, the fit in which they allowed the
linewidth to vary was not a significant improvement over that in which
they pinned the width at $\sigma = 100$~eV. Thus, their line width
constraints are also consistent with the majority of the spectra
described by Fig.~\ref{fig_widvsshift}.

\section{Other signatures of a fast outflow}
\label{sect_beyondka}

As discussed in Section~\ref{sect_intro}, blueshifted Fe~K$\alpha$
absorption is the clearest signature of highly ionized
outflowing gas and has therefore been the central topic of this paper.
However, as mentioned in Section~\ref{sect_example}, narrow
blueshifted Fe~K$\alpha$ absorption is neither a necessary consequence
nor the only signature
of a highly ionized flow. 
Thus if such outflows are common in the AGN population it is
important to consider other effects they may have on X-ray spectra. 
In this section, we briefly describe some of these effects and their
possible implications with reference to our model spectra.

\subsection{Blueshifted absorption lines other than Fe~K$\alpha$}

Although the Fe~$\alpha$ feature is typically the most prominent
absorption line, K$\alpha$ absorption by lighter elements is present
in many of the spectra computed from our grid of models, particularly
those with high $\dot{M}$-values.
The strongest of these K$\alpha$ features are those of S, Si and O  -- 
the absorption EWs for these lines are up to 10, 10 and 3~eV respectively.
Their strengths 
correlate closely with each other and more loosely with
the Fe~K$\alpha$ feature.

At energies harder than Fe K$\alpha$, absorption by H-/He-like Ni
K$\alpha$ and by Fe K$\beta$/K$\gamma$ occur. 
Individually, these features are all weaker than Fe~K$\alpha$ but 
they are often
blended together producing fairly significant absorption which can
extend up to around 10~keV.

These additional absorption lines can provide
both confirmation of outflows detected via Fe~K$\alpha$ and
supplementary constraints on the flow properties. However, their
reliability is limited since (i) in our models, they are always weaker
and thus harder to detect than Fe~K$\alpha$ and (ii) with the
exception of the hard energy Ni or Fe~K$\beta$ lines, their
identifications are less clear since there are more potential atomic
transitions at soft energy. Thus, as diagnostics for highly ionized
flows, these lines are always of subordinate value to Fe~K$\alpha$.

\subsection{Emission lines and P-Cygni profiles}

As pointed out in the reference spectra shown in
Section~\ref{sect_example}, the blueshifted absorption lines are often
accompanied by broad emission which is centred close to the line rest
energy. Since these emission features are formed as a result of line
scattering and recombination emission in the outflow, their strength
strongly depends on the density of the wind; thus although they
are fairly weak in the spectra from the example
model, they become more significant in our models with higher
mass-loss rates. Reliable predictions for these phenomena in
non-spherical outflows requires multi-dimensional radiative transfer
and thus our methods are particularly well-suited to describing them.

\begin{figure*}
\epsfig{file=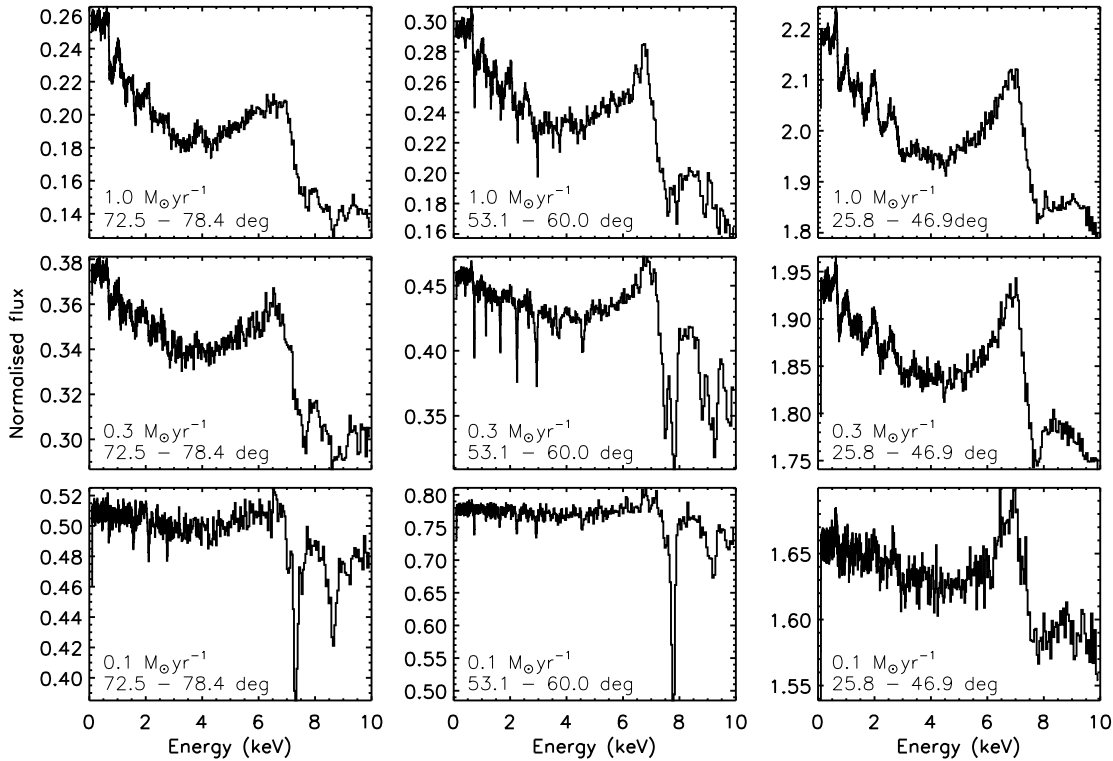, width=15cm}
\caption{
Sample spectra computed for different viewing angle bins (left to
right) for models with differing mass-loss rates (top to bottom).
Except for $\dot{M}$, all model parameters are fixed at those of the
example model. The plotted spectra are all normalised to the input
primary power-law spectrum.
}
\label{fig_emiss}
\end{figure*}

To illustrate the types of line profile, Fig.~\ref{fig_emiss} shows a
montage of spectra from models with relatively high mass-loss rates.
There is a wide range of K$\alpha$ profile types in these models
and their emission EWs, measured relative to a power-law continuum
fit, can be in excess of 200~eV; 
{this is comparable to the typical 
broad K$\alpha$ EWs measured in a sample of AGN by \citet{nandra07}.}

At low inclination angles,
our model Fe K$\alpha$ profiles have a near-P-Cygni
character: they shows some blueshifted absorption and broad, redshifted
emission, the strength of which grows with $\dot{M}$. Such profiles
are qualitatively similar to those observed in 
some Seyfert 1 galaxies \citep{done07}.

At intermediate angles, when the line-of-sight is close to looking
down the wind cone, the narrow absorption lines are strongest but
are accompanied by moderately strong,
emission lines with fairly extended red wings. We note that
the narrow absorption lines become less prominent at very high
$\dot{M}$, a consequence of the increased contribution of scattered
radiation in the spectra.

For the highest inclinations and $\dot{M}$-values, the Fe~K$\alpha$
line is almost pure emission, peaking at the rest energy of the line
and having a long red tail. The redward extension of the
line emission is not a consequence of gravitational redshift but
arises from electron scattering in the outflow. 

{ 
Although somewhat more diverse,
our red-skewed emission line profile shapes are qualitatively similar to
those obtained via the same physical mechanism in spherical outflow 
models (\citealt{laurent07}; see also \citealt{laming04}) while they
are significantly broader than those computed by \citet{rozanska08} 
for Compton scattering in irradiated accretion disks. 
The largest Fe K$\alpha$ emission 
EWs found by \citet{laming04} (\simgt 4keV) substantially exceed those
found in any of our models but we note that the large EWs in their
models mostly arise from significantly lower ionization stages than are
present in our models.
}

These emission line properties show that
highly ionized outflow may affect the Fe~K region beyond
imprinting narrow, blueshifted absorption lines. 
This may have
consequences for study of the Fe~K fluorescent emission which
originates in AGN accretion disks 
(e.g. \citealt{fabian89,nandra97,fabian00,jmiller07,nandra07}),  
since it may
contaminate the disk emission and/or lead to an apparently
multi-component emission line (see e.g. \citealt{oneill07}).
In this context, the red wings predicted for the emission lines may be
of particular relevance in view of the potential for confusion with
the effects of gravitational redshift -- however, more
sophisticated 3D 
models going beyond the smooth-flow assumption of
our parameterised wind models must be examined before such a
possibility can be considered in much greater detail.

\subsection{Spectral curvature}

Absorption by outflowing material has been discussed as a possible
explanation for the so-called soft-excess in X-ray spectra \citep{gierlinski04,gierlinski06,middleton07,schurch07,schurch08}. In this picture, the decrease in flux
above $\sim 1$~keV is attributed to absorption by light or
intermediate mass elements. 

Such absorption does occur in our models, particularly for
high $\dot{M}$-values (see Fig.~\ref{fig_emiss}) however the scale of
the effect is too small in the models presented here: the
typical observed soft excess requires a drop in flux of nearly a
factor of two between about 1 and 2 keV
\citep{middleton07}. Furthermore, \citet{schurch08} have argued that
very large, relativistic velocities are required for the absorption
model to work since the observed soft excess spectra are very
smooth. Our current models support their conclusions since all cases
in which significant spectral curvature arises are accompanied by
discreet features.

Further consideration must await extension of our calculations to
low-ionization states since more significant absorption at relatively
soft energies can then be expected.

\section{Detailed comparison with Mrk 766}
\label{sect_mrk766}

One of the AGN whose X-ray spectrum provided some of the initial
motivation for this work is Mrk\,766 \citep{miller07, turner07}.
As well as the possible outflow absorption lines already mentioned in
section\,\ref{sect_obs}, \citet{miller07} suggested that
the extremely hard 2 -- 10~keV spectrum in the low flux state, coupled
with the strong Fe edge and weak Fe emission line, might be a signature
of either ionised absorption or ionised reflection, possibly from an
obscuring wind.  Such a picture provides an alternative explanation to
the hypothesis of ``relativistic blurring'' that has been invoked to
explain the apparent ``red wing'' at energies below 7~keV in 
the mean X-ray spectra of AGN, 
including Mrk\,766 \citep[e.g.][]{nandra07}.

\begin{figure}
\epsfig{file=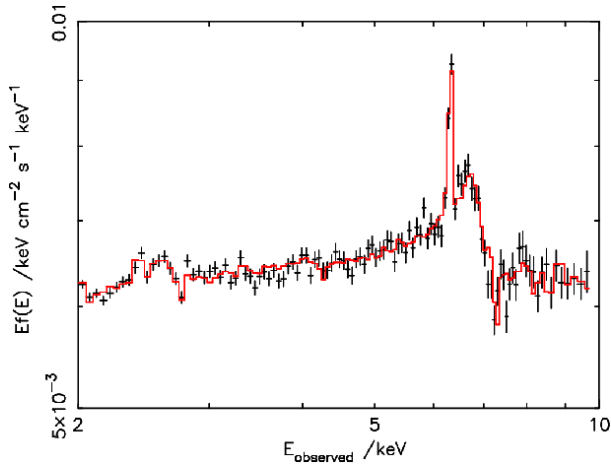, width=8cm}
\caption{
The best-fitting wind model fit to the mean 2 -- 10~keV
spectrum of Mrk\,766, plotted in units of Ef(E),
showing model and ``unfolded'' data.
}
\label{mrk766}
\end{figure}

We now make a direct comparison between the wind model in this paper and
the mean spectrum of Mrk\,766 obtained by summing the {\em XMM-Newton}
{\sc epic pn} data from the 2000, 2001 and 2006 observations. 
Observation IDs were 0096020101, 0096020101 and 0304030[1-7]01
and data were processed as described by \citep{miller07}.  As in that work,
energy bins of width equal to the energy-dependent HWHM were adopted.
The model was fitted to the data by first creating a 
multiplicative table (``mtable'') covering a range of parameters that could
be used with {\sc xspec} \citep{arnaud96}.  
{
The models used for this table were those described in Sections 5.1,
5.2 and 5.3 making up a regular, four-dimensional grid covering three of our
most important wind parameters
($r_{\mbox{\scriptsize min}}$, $d$ and $\dot{M}$) and the
viewing-angle ($\theta$). The other wind parameters were not varied
in the fit.
}
The table was used to multiply 
an assumed intrinsic power-law whose photon index was a free parameter
in the fit: in principle the fitting is valid only for the
choice of power-law index used to create the model, but in practice we find
that there is little variation between models whose power-law indices
differ, and that the variation that is seen is to some extent degenerate
with the effects caused by varying other wind parameters (see Section~\ref{sect_grid}).  We expect also
some amount of spectral curvature caused by the ionised absorption previously
identified in Mrk\,766, so we add a low column of ionised absorption
calculated by {\sc xstar} \citep{kallman04} and we model the narrow
6.4~keV\,Fe emission line as arising in mildly ionised reflection
calculated by the {\sc reflionx} model \citep{ross05} with 
no relativistic blurring.  
Owing to computing time limitations the model grid was necessarily
rather coarse in parameter space, so to check the fit of the best-fitting
model, interpolated between models, we then recalculated the model with
precisely the parameter values indicated by fitting.
The wind best-fit parameter were 
$r_{\rm min}=385$\,r$_{\rm g}$,
$d = 1.6 r_{\rm min}$,
$\dot{M} = 0.4 $\,M$_\odot$\,yr$^{-1}$ and
$\cos\theta=0.77$, while the best-fit photon index was $\Gamma=2.03$
and absorber parameters were N$_{\rm H} = 5.3 \times 10^{21}$\,cm$^{-2}$
and ionization parameter $\xi = 40$\,erg\,cm\,s$^{-1}$.
These parameters yielded a goodness-of-fit $\chi^2 = 145$ for 108
degrees of freedom (no component of systematic error is assumed).
The fit
to the spectrum in the 2 -- 10~keV range is shown in Fig.\,\ref{mrk766},
where we plot the model spectrum at the {\em XMM-Newton} resolution together
with ``unfolded'' data points.

The fit of the wind model to the data does indicate that this may well
provide a good explanation of the shape of X-ray spectra in AGN such
as Mrk\,766.  With the exception of the narrow Fe emission line,
most of the spectral shape in the Fe regime is provided by the wind
model, with the reflection component contributing only 3\,per cent
of the total 2 -- 10~keV energy flux and only 4\,per cent of the flux
density in the continuum around Fe.
It is unlikely that the current model can explain the
full spectral variability seen in this AGN's various flux states: the
hard spectrum in the low state is most likely caused by moderately
ionised material, $\xi \simeq 100$\,erg\,cm\,s$^{-1}$ \citep{miller07},
which is not yet treated in the wind model.  We postpone to future work
fits to the X-ray spectra of this and other AGN 
with enhanced, lower-ionization, wind models.

\section{Conclusions and future work}
\label{sect_conc}

We have used a simply-parameterised model to construct 
a variety of possible AGN outflow geometries and computed spectra for
various lines-of-sight using a multi-dimensional 
Monte Carlo radiation transport code.

Our results show that narrow blueshifted Fe~K$\alpha$ 
absorption lines are a natural consequence of highly ionized
flows and that their formation depends strongly 
on the flow inclination angles relative to the
observer's line-of-sight. Typically, our models suggest that 
moderately strong features K$\alpha$ absorption features (\simgt
40~eV) will manifest for a significant minority of possible viewing
angles.

Although blueshifted absorption lines are the clearest signature of an
outflow, our calculations have illustrated a number of other
spectroscopic signatures which may be relevant to the interpretation
of X-ray data in the context of AGN outflows. In particular, electron
scattering, line scattering and recombination emission can all affect
the Fe~K region. For suitable wind conditions and
lines-of-sight, these processes can 
lead to significant net emission in K$\alpha$ and can make
a strong red electron-scattering line wing. We have illustrated that
these emission line profiles look qualitatively similar to observed
spectral features (e.g. in Mrk~766) suggesting that outflows may have
a much more general role in the formation of AGN X-ray spectra than
simply imprinting absorption lines. At the very least,
emission/scattering features such as those found in our models must be
considered as a potential source of contamination for studies of other
sources of Fe~K emission.

{
Considerable further work is required to more fully understand the
role of outflows in the spectral formation of AGN. 
The primary limitation of this study is the restriction to reliable
modelling of only K-shell ions. 
K-shell transitions occurring in lower ions, as may be present in
less ionized wind regions, will have a qualitative effect on the Fe~K
region -- e.g. while the K$\alpha$ line of H- and He-like ions predominantly
scatter incoming photons, the Li- and Be-like ions can destroy them
with significant probability.   
Fortunately, the restriction to K-shell ions is not a
fundamental limitation of the methods used but was made
primarily in the interests of computational expediency for this
study of Fe~K shell absorption features. 
} The next
step will be to extend the physics of the code to incorporate L-shell ions
so that lower ionization states may be probed. In particular, this
will allow the investigation of outflow geometries with larger
$r_{\mbox{\scriptsize max}}$ and flows with a clumpy structure.
Furthermore, in this study  we
have only considered the simplest X-ray source location and have
neglected any interplay between different components in the AGN
environment. In particular, we have not considered how emission,
absorption 
and reflection by an accretion disk will affect the spectral
signatures of outflow, and vice versa. Ultimately, it will be
desirable to construct physically motivated models for complete
disk/outflow systems and use these to interpret the increasing wealth
of X-ray observations. 

\section*{Acknowledgments}
 
S. A. Sim acknowledges V. Wild for very useful discussions.
T. J. Turner acknowledges NASA Grant ADP03-0000-00006.
{ 
We thank the anonymous referee for constructive suggestions.
}

\bibliographystyle{mn2e}
\bibliography{snoc}

\label{lastpage}

\end{document}